\documentclass[a4paper,aps,onecolumn,showpacs,showkeys,nofootinbib,preprintnumbers,superscriptaddress,amsmath,amssymb,amsfonts]{revtex4}
\usepackage{graphicx}
\usepackage{bm,natbib}
\usepackage{amsmath}
\usepackage[english]{babel}
\usepackage[T1]{fontenc}
\usepackage{amssymb}
\usepackage{amsfonts}
\usepackage{epsfig}
\usepackage{colordvi}
\usepackage{psfrag}
\usepackage{color}
\usepackage{ mathrsfs }
\newcommand{\Lagr}{\mathcal{L}}
\newcommand{\G}{\mathcal{G}}
\newcommand{\al}{\alpha}
\newcommand{\be}{\beta}
\usepackage{dcolumn}
\usepackage{multirow}
\usepackage{hyperref}
\usepackage{epstopdf}
\usepackage{bm}
\usepackage{times}

\hypersetup{
  colorlinks=true,        
  linkcolor=blue,         
  citecolor=magenta,      
}

\begin{document}
\title{$f(\G)$ Noether  cosmology}


\author{Francesco Bajardi}
\affiliation{Universit\'a di Napoli "Federico II", Compl.~Univ.~di Monte S.~Angelo, Edificio G, Via Cinthia, I-80126, Napoli, Italy}
\affiliation{INFN Sezione  di Napoli, Compl. Univ. di
Monte S. Angelo, Edificio G, Via Cinthia, I-80126, Napoli, Italy.}

\author{Salvatore Capozziello}
\affiliation{Universit\'a di Napoli "Federico II", Compl.~Univ.~di Monte S.~Angelo, Edificio G, Via Cinthia, I-80126, Napoli, Italy}
\affiliation{INFN Sezione  di Napoli, Compl. Univ. di
Monte S. Angelo, Edificio G, Via Cinthia, I-80126, Napoli, Italy.}
\affiliation{Gran Sasso Science Institute, viale F. Crispi 7, I-67100, L' Aquila, Italy.}
\affiliation{Tomsk State Pedagogical University, ul. Kievskaya, 60, 634061 Tomsk, Russia.}
\date{\today}

\begin{abstract}
We develop  the $n$-dimensional cosmology  for  $f(\G)$ gravity, where $\G$ is the \emph{Gauss-Bonnet} topological invariant. Specifically,  by the so-called  Noether Symmetry Approach, we   select  $f(\G)\simeq \G^k$ power-law models where $k$ is a real number.  In particular,  the case   $k = 1/2$ for $n=4$ results   equivalent to  General Relativity showing that we do not need to impose the action $R+f(\G)$  to reproduce the Einstein theory.    As a further result,  de Sitter solutions are  recovered in the case where $f(\G)$ is non-minimally coupled to a scalar field. This means that issues like inflation and dark energy can be addressed in this framework. Finally,  we develop the Hamiltonian formalism for the related minisuperspace and discuss the quantum cosmology for this model.
\end{abstract}

\pacs{98.80.-k, 95.35.+d, 95.36.+x}
\keywords{Alternative theories of gravity; cosmology; exact solutions.}

\maketitle

\section{Introduction}
Despite the successes and  probes of General Relativity (GR), it presents  issues  at IR  and UV scales pointing out that it is not the final theory of gravity   \cite{weinb, Gronwald:1995em}. Clearly there are problems with quantization of spacetime geometry (the lack of a final Quantum Gravity Theory) and with large scale structure (the unknown dark side to fit astrophysical and cosmological dynamics).

In this context, modified theories of gravity (obtained by extending or changing  the Hilbert-Einstein action) could be suitable to fix  Dark Energy and  Dark Matter issues emerging along the cosmic history.  Basically, the philosophy consists in considering extended/modified gravitational  Lagrangians where extra-terms in the field equations could  play the role of  the "Dark" components  and explain the expansion of the universe and the large scale structure.  Dark Matter and Dark Energy, in fact, represent a controversial problem in cosmology and astrophysics, since they are supposed to cover almost  95\% of the universe content  but have never been observed at fundamental scales even if they manifest their effects at large scales.

Nevertheless, extending/modifying  GR allows to overcome several other issues, see e.g.  \cite{Capozz10, Capozz8}. In particular, they provide new polarization modes for  gravitational waves \cite{Capozziello:2019klx}, are capable of describing the fundamental plane of galaxies \cite{Jovanovic:2016ava, Borka}, can fit  Dark Energy dynamics \cite{Sharif:2012ig, Nojiri:2006ri, Bahamonde:2018miw, Capozziello:2002rd, Odintsov1,Odintsov2},  can address astrophysical structures through corrections to the Newtonian potential \cite{Capozziello:2012ie}. From another point of view, modified theories may better adapt to the QFT formalism for several reasons: it is possible to extend the action in order to construct super-renormalizable theories \cite{Modesto:2017hzl} or to build up  effective  theories towards  quantum gravity \cite{Horava:2009uw}.

Among  extended theories of gravity, $f(R)$ gravity  is a straightforward extension where the hypothesis of linearity of the Ricci scalar $R$ into the Hilbert-Einstein action is relaxed 
\cite{Capozz1,  Capozz7, Capozz4, Capozz5, De Felice, Starobinsky1,Starobinsky2, Capozziello:2019cav}.   

However also non-minimal   couplings  as in  the \emph{Brans-Dicke} theory as well as  higher-order curvature invariants like $f(R, \Box R, \Box^2 R...\Box^k R)$ or $f(R, R^{\mu \nu} R_{\mu \nu}, R^{\mu \nu p \sigma} R_{\mu \nu p \sigma})$ can be considered in this program of extending GR \cite{Capozz1, Odintsov1, De Felice}. 

Another point of view approaches gravity as a theory of the translational group where affinities play a major role  \cite{Capozz3, ft, Pereira}. It is  the so called \emph{Teleparallel Equivalent General Relativity}  where the antisymmetric part of the  connection is taken into account, the so called Weitzenb\"ock connection,  constructed on tetrads \cite{Hamid, Pereira3, Pereira4}.  In this picture,    spacetime dynamics is given by torsion instead of curvature and the Equivalence Principle is not necessary to fix causal and geodesic structure. 

Finally, also non-metricity can be considered in the debate of define the physical variables of gravity and then curvature, torsion and non-metricity or, alternatively, metric, tetrads and non-metricity can give the geometric description of the gravitational field \cite{Lavinia}. 

In this perspective, a peculiar role is played by the topological  invariants which intervene in the formulation of any quantum field theory  on curved space. Specifically, the so called Gauss-Bonnet  (GB) topological invariant has a crucial role in trace anomaly and in the regularization of the theory at least at one-loop level \cite{Birrell}. This invariant can have an important role also in cosmology as pointed out in 
  \cite{Odintsov3,Odintsov4}. It emerges in gravitational actions containing second-order curvature invariants like 
 \begin{equation}
 S = \int \sqrt{-g} f(R,R^{\mu \nu} R_{\mu \nu}, R^{\mu \nu p \sigma} R_{\mu \nu p \sigma}) d^4 x\,,
 \end{equation} 
 see for example \cite{SO}.  Here, the curvature terms can be combined as 
 \begin{equation}
 \label{GBcovariant}
  \mathcal{G} = R^2 - 4R_{\mu \nu} R^{\mu \nu} + R_{\mu \nu p \sigma} R^{\mu \nu p \sigma}\,,
  \end{equation}
 which is the GB invariant acting as a constraint among the second-order curvature terms. According to this consideration, a theory like $f(R,\mathcal{G})$ can exhaust all the degrees of freedom related to the second-order curvature invariants which is  dynamically equivalent to a theory with two scalar fields
 \cite{Paolella}.

  However, in (3+1)-dimensions, it is well known that an action like $S = \int \sqrt{-g} \G \, d^4x$ is trivial, while  it is not so   in (4+1)-dimensions (or more).
 This result is directly linked to the GB theorem, which states that the integral of the GB term over the manifold is the Euler characteristic of the manifold, \emph{i.e} a topological invariant. 
 
 Despite of this result, considering any function $f(\G)\neq \G$ can be  mathematically and physically relevant, also in 4D,  for the following reasons. In general, $f(R,\G)$ gravity is taken into account to recover GR, in a given limit, assuming $f(R,\G)=R+f(\G)$. Observational and theoretical constraints have been obtained also for other forms of $f(R,\G)$ \cite{Santos,Ben} but pure $f(\G)$ theories are not, in general,  considered because GR seems excluded. 
 
 In this paper,  we  deal with actions  depending only on the topological surface term. From a mathematical  point of view,    the related dynamics is simpler and often  analytically solvable. 
  
 Besides this technical point, it is worth stressing that, in homogeneous and isotropic cosmology,  terms containing squared Ricci and Riemann tensors contribute dynamically as the squared Ricci scalar in the GB invariant. This is particularly evident for  the cosmological scale factor evolving as   exponential function or   power-law functions.  According to this observation,  as soon as  $f(\G) \sim \G^{\frac{1}{2}}$ holds, a GR-like behavior is expected. This means that, by  a theory containing only  topological terms,   GR  can be, in principle, recovered   without inserting by hands  the Hilbert-Einstein term in the gravitational action.  In other words, starting from a theory  regular and consistent with quantum considerations \cite{Birrell}, one could recover GR results at IR scales avoiding some pathologies.
 
A detailed treatment of the GB action in  spherically symmetric configuration is reported in \cite{Baj}. 

As a general remark, considering theories as $R+f(\G)$ and $f(R)+f(\G)$ is particularly useful in high energy regimes of gravity. Specifically, the introduction of $\G$ leads to improve  the inflationary scenario, where  two acceleration phases can be  led by $R$ and $\G$ respectively giving rise to  a $R-$dominated phase and a $\G-$dominated phase. Being $\G$ dominant in stronger curvature regimes, its contribution, through a non-linear function $f(\G)$, rules  the universe behavior at very  early stages  extending the Starobinsky model \cite{Paolella}. Both $f(R)$ and $f(\G)$ fields can cooperate to the slow rolling phase with a  behavior  depending on the strength of the corresponding coupling constant. In general, the contributions of $f(R)$ and $f(\G)$ give rise to a potential, whose minima can be separated by a barrier, representing a double inflationary scenario where the Gauss-Bonnet term dominates at very early epochs and  the Ricci scalar at moderate early epochs.  Finally, realistic scenarios  converge towards standard GR. On the other hand, $f(\G)$ terms contribute to late accelerated expansion as discussed in \cite{Odintsov3,Odintsov4} and theories like $f(R,\G)$ satisfy the Solar System constraints  \cite{DeLaurentis:2013ska}. 
 
 This work is  focused on  $f(\G)$ cosmology studied by the \emph{Noether Symmetry Approach} \cite{Capozziello2}. The main result is that the existence of symmetries selects models of the form $f(\G)=\G^k$, with $k$ any real number. Then, for $k=1/2$, GR is recovered as one  of the models  allowed by symmetries.
 
As pointed out in \cite{Montesinos:2020pxv, Montesinos:2018zia, Montesinos:2017epa}, the search for symmetries in modified theories of gravity plays a fundamental role in order to get suitable equations of motion through a selection criterion  motivated by physical reasons.
 
 In this framework, as shown in \cite{Capozz7, Capozz5, Capozz3, Capozz8, Capozziello2, Capozz13}, thanks to Lagrange multipliers, one can find the cosmological point-like Lagrangians to develop the Noether approach.  Adopting a  \emph{Friedmann-Robertson-Walker} (FRW) metric in 4-dimensions, the Lagrange multiplier for $f(\G)$ results
  \begin{equation}
  \label{lagrange}
   \G = 24 \left(\frac{\ddot{a}\dot{a}^2}{a^3}\right)\,.
   \end{equation}
   It is easy to see that  it comes from a 4-divergence 
    \begin{equation}
     \sqrt{-g} \, \G = 8 \frac{d}{dt}(\dot{a}^3)\,,
     \end{equation}
and hence, when integrated, it  gives only  a trivial contribution. In other words,  when we consider the extension $f(\G)$, a straightforward integration by parts of the Lagrange multiplier allows to write the non-vanishing point-like canonical  Lagrangian as a function of the scale factor $a$,  the field $\G$ and their first  derivatives. We have  a configuration space $\mathcal{Q}\equiv\{ a,\G \}$ equipped with tangent space $T\mathcal{Q}\equiv\{ a,\dot{a},\G, \dot{\G}\}$. This is a 2-dimensional \emph{minisuperspace} which can be easily quantized in view of  quantum cosmology considerations.
    
The layout of the paper is the following.     In  Sec. \ref{GB cosmo}), we introduce the main features of the GB gravity and cosmology in $n$ dimensions. Sec. \ref{noether} is devoted  to the  Noether Symmetry Approach  for $f(\G)$ cosmology.  In Sec. \ref{soluz},  we find  symmetries for GB cosmology, showing that we can recover GR and find exact  cosmological solutions.  Secs. \ref{branse} and \ref{sum} are respectively devoted to the coupling between modified GB action and a scalar field, and models given by a  sum of GB functions. Quantum cosmology considerations are developed in Sec. \ref{Hamiltonian}, where we find the Wave Function of the Universe for the minisuperspace $T\mathcal{Q}\equiv\{ a,\dot{a},\G, \dot{\G}\}$.
Discussion and conclusions are reported in Sec. \ref{concl}.

\section{Gauss-Bonnet Cosmology}
\label{GB cosmo}
Let us now discuss some basic results of GB gravity and cosmology. First of all, we introduce the GB topological invariant  $\mathcal{G}$. In $n$-dimensions, assuming gravity as a gauge theory of the local Lorentz group on the tangent bundle, the GB term is:
\begin{equation}
\G = \epsilon_{a_1,a_2,a_3....a_n} R^{a_1,a_2} \wedge R^{a_3,a_4} \wedge e^{a_5} \wedge ... \wedge e^{a_n},
\end{equation} 
being $R^{a_i,a_j}$ the two form curvature, $e^k$ the set of zero forms defining the basis and $\epsilon_{a_1,a_2,a_3....a_n}$ the Levi-Civita symbol. This GB term  is part of the $n$-dimensional Lovelock Lagrangian \cite{Lovelock, Montesinos:2018ujm} which,  in four dimensions, can be expressed as:
\begin{equation}
L^{(4)} =\epsilon_{abcd} \left[ \alpha_2 R^{ab} \wedge R^{cd} + \alpha_1 R^{ab} \wedge e^c \wedge e^d + \alpha_0 e^a \wedge  e^a \wedge  e^b \wedge  e^c \wedge  e^d\right] \;,
\end{equation} 
where  the first term  is the GB invariant, the second the Ricci scalar and the third the cosmological constant. Even though the Gauss Bonnet term naturally emerges in the Lovelock gravity under the gauge formalism,  we will deal with the covariant representation of $\G$, which is given by Eq.\eqref{GBcovariant}. Here, we consider a general analytic function of $\G$ and the action 

\begin{equation}
S = \int \sqrt{-g} f(\G) \; d^nx \,.
\end{equation}
Varying it with respect to the metric, we find the following field equations 
\begin{eqnarray}
&&2R \nabla_\mu \nabla_\nu f'(\G) - 2g_{\mu \nu} R \Box f'(\G) - 4R^\lambda_\mu \nabla_\lambda \nabla_\nu f'(\G) + 4R_{\mu \nu} \Box f'(\G) \nonumber
\\
&&+ 4 g_{\mu \nu} R^{p \sigma} \nabla_p \nabla_\sigma f'(\G) + 4 R_{\mu \nu p \sigma} \nabla^p \nabla^\sigma f'(\G) + \frac{1}{2} g_{\mu \nu}[f(\G) - \G f'(\G)] = T_{\mu \nu} \;,
\label{field equations}
\end{eqnarray}
where $\Box$ is the $n$-dimensional D'Alembert operator ($\Box = g_{\mu \nu} \nabla^\mu \nabla^\nu$) and the prime indicates the derivative with respect to $\G$. $T_{\mu \nu} $ stands for the energy-momentum tensor of  matter and, for simplicity, we used physical units ($\hbar = c = k_B = 8\pi G = 1$). 
In \cite{DeLaurentis:2013ska, SantosdaCosta:2018ovq, Andrew:2007xa, Ivanov:2011vy}, one can found the generalization of this action to the case $f(R,\G)$ in 4-dimensions.

In order to obtain the form of the GB scalar in cosmology,  we have to  calculate the $n$-dimensional Riemann tensor, Ricci tensor and Ricci scalar in  FRW metric. We choose for the interval 
\begin{equation}
ds^2 = dt^2 - a(t)^2 \delta_{ij} dx^i dx^j \,,
\end{equation}
where the index $i,j$ label all the spatial dimensions and run from $1$ to $n$. We assume the spatially flat case. The non-null curvature components are:
\begin{eqnarray}
&&R = -2(n-1) \frac{\ddot{a}}{a} - (n-2) (n-1) \frac{\dot{a}^2}{a^2}\,, \qquad R_{00} = (n-1) \frac{\ddot{a}}{a}\,, \qquad R_{ij} = \left[(n-2) \dot{a}^2 + a \ddot{a} \right] \delta_{ij}\,, \nonumber
\\
&& R_{0i0j} = a \ddot{a} \delta_{ij}\,, \qquad R_{ijm \ell} =a^2 \dot{a}^2 \delta_{im} \delta_{j \ell}\,.
\end{eqnarray}
By properly contracting the above quantities, the n-dimensional GB term turns out to be
\begin{equation}
\G = \frac{(n-3)(n-2)(n-1) \left[(n-4) \dot{a}^4 + 4 a \dot{a}^2 \ddot{a}\right]}{a^4} \equiv p(n) \frac{\left[(n-4) \dot{a}^4 + 4 a \dot{a}^2 \ddot{a}\right]}{a^4},
\label{nDG}
\end{equation}
with $p(n) = (n-1)(n-2)(n-3)$. As we can see, in less than four dimensions it vanishes regardless of the value of the scale factor, while in 4-dimensions, it turns into a topological surface term of the form given in Eq.\eqref{lagrange}.

Dynamics can be derived both starting from field Eqs. \eqref{field equations} or from the Euler-Lagrange equations derived from a point-like  Lagrangian. Because of our further considerations related to the Noether Theorem,  let us construct the point-like  Lagrangian.  It can be found thanks to the Lagrange multipliers method, with constraint \eqref{nDG}, as follows:
\begin{equation}
S = \int \left[\sqrt{-g} f(\mathcal{G}) - \lambda \left\{ \mathcal{G} - \frac{(n-3)(n-2)(n-1) \left[(n-4) \dot{a}^4 + 4 a \dot{a}^2 \ddot{a}\right]}{a^4} \right\} + \Lagr_m \right] d^nx \;,
\end{equation}
being $\Lagr_m$ the matter Lagrangian. Considering the cosmological volume element in $n$-dimensions,  the action can be written as
\begin{equation}
S =2 \pi^2 \int \left[a^{n-1} f(\mathcal{G}) - \lambda \left\{ \mathcal{G} - \frac{(n-3)(n-2)(n-1) \left[(n-4) \dot{a}^4 + 4 a \dot{a}^2 \ddot{a}\right]}{a^4} \right\} + \Lagr_m \right] d^nx \;.
\label{azione con lambda}
\end{equation}
By varying the action with respect to $\mathcal{G}$, we are able to find $\lambda$:
\begin{equation}
\delta S = \frac{\partial S}{\partial \mathcal{G}} \delta \mathcal{G} = a^{n-1} f'(\mathcal{G}) - \lambda = 0 \;\;\;\;\;\; \lambda = a^{n-1} f'(\mathcal{G}) \;.
\end{equation}
Replacing in Eq. \eqref{azione con lambda} and integrating out the second derivative, the Lagrangian finally takes the form:
\begin{equation}
\Lagr = \frac{1}{3} a^{n-5} \left[(4 - n) p(n) \dot{a}^4 f'(\G)+ 3 a^4 [f(\G) - \G f'(\G)] - 4 a p(n) \dot{a}^3 \dot{\G} f''(\G) \right] + \Lagr_m
\label{lagrangiana nD}
\end{equation}
The dynamical system is given by  the  two Euler-Lagrange equations coming from Lagrangian \eqref{lagrangiana nD}, with respect to the scale factor $a$ and the GB scalar $\G$. The system is completed by  the Energy condition $E_\Lagr = \displaystyle \left(\dot{a} \partial_{\dot{a}} + \dot{\G} \partial_{\dot{\G}}- 1 \right) \Lagr = 0$.  Finally we have
\begin{equation}
\begin{cases}
 -(n-4)(n-5) p(n) \dot{a}^4 f'(\G) - (n-1) a^4 [f(\G) - \G f'(\G)] - 4 (n-4) p(n) a \dot{a}^2 [\ddot{a} f'(\G)+ \dot{a} \dot{\G} f'(\G)] + 
\\
\displaystyle - 4 a^2 p(n) \dot{a} [2 \dot{\G} \ddot{a} f''(\G) + \dot{a} \ddot{\G} f''(\G) + \dot{a} \dot{\G}^2 f'''(\G)]=0 
\\
\\
 \G =p(n) \frac{\left[(n-4) \dot{a}^4 + 4 a \dot{a}^2 \ddot{a}\right]}{a^4}
\\
\\
 (4 - n) p(n) \dot{a}^4 f'(\G) - a^4 [f(\G) - \G f'(\G)] - 4 a p(n) \dot{a}^3 \dot{\G} f''(\G) = 0.
\label{EL3}
\end{cases}
\end{equation}
It is worth noticing  that the equation for $\G$ provides exactly the cosmological constraint on  the GB scalar \eqref{nDG}.  It is impossible to solve the above equations without  selecting the form of the $f(\G)$ function. In order to do this, we adopt the Noether Symmetry  Approach by which one can  select reliable models  according to the existence of  symmetries. The approach is also physically motivated because symmetries correspond to conservation laws. 

\section{The Noether Symmetry Approach}
\label{noether}
In this section we sketch the  Noether Symmetry Approach \cite{Capozziello2}, that we will apply in the next section, to   the previous cosmological point-like Lagrangian. Let us consider the following transformations which leave the Euler-Lagrange equations invariant with respect to a change of coordinates: 
\begin{equation}
\begin{cases}
\Lagr(t,q^i \dot{q}^i) \to \Lagr (\overline{t}, \overline{q}^i, \dot{\overline{q}}^i)
\\
\overline{t} = t + \epsilon \xi(t,q^i) + O(\epsilon^2)
\\
\overline{q}^i = q^i + \epsilon \eta^i(t,q^i) + O(\epsilon^2) \;.
\end{cases}
\label{trasformazioni}
\end{equation}
In order to find the  generator of  transformations,  we need to find the transformation law of the first derivative since, being the time involved in the transformation, the quantity $\dot{\overline{q}}^i$ does not trivially correspond to $\displaystyle \frac{d \overline{q}^i}{d t}$. For the first derivative,  we have:
\begin{equation}
\dot{\overline{q}}^i = \frac{d \overline{q}^i}{d \overline{t}} = \frac{d q^i + \epsilon d\eta^i(t,q^i)}{dt + \epsilon d \xi(t, q^i)} = \displaystyle \frac{\frac{d q^i}{dt} + \epsilon \frac{d\eta^i(t,q^i)}{dt}}{1 + \epsilon \frac{d \xi(t, q^i)}{dt}} = \left(\frac{d q^i}{dt} + \epsilon \frac{d\eta^i(t,q^i)}{dt}\right) \left(1+ \epsilon \frac{d \xi(t, q^i)}{dt}\right)^{-1} \;.
\end{equation}
which, up to the first order, takes the form
\begin{equation}
\dot{\overline{q}}^i = \left(\frac{d q^i}{dt} + \epsilon \frac{d\eta^i(t,q^i)}{dt}\right) \left(1 - \epsilon \frac{d \xi(t, q^i)}{dt}\right) + O(\epsilon^2) =
\nonumber
\end{equation}
\begin{equation}
= \frac{d q^i}{dt} + \epsilon \left[\frac{d\eta^i(t,q^i)}{dt} - \frac{d q^i}{dt} \frac{d \xi(t, q^i)}{dt}\right] + O(\epsilon^2) \sim \dot{q}^i + \epsilon \left[ \dot{\eta}^i - \dot{q}^i \dot{\xi} \right] \;.
\end{equation}
Let us now define $\eta^{i \; [1]} = \dot{\eta}^i - \dot{q}^i \dot{\xi}$  so that we have:
\begin{equation}
\begin{cases}
\Lagr(t,q^i \dot{q}^i) \to \Lagr (\overline{t}, \overline{q}^i, \dot{\overline{q}}^i)
\\
\overline{t} = t + \epsilon \xi(t,q^i) + O(\epsilon^2)
\\
\overline{q}^i = q^i + \epsilon \eta^i(t,q^i) + O(\epsilon^2)
\\
\dot{\overline{q}}^i = \dot{q}^i + \epsilon \eta^{i \; [1]}
\end{cases}
\label{trasformazione coordinate}
\end{equation}
and finally the generator of  transformation has the form
\begin{equation}
X^{[1]} = \xi \frac{\partial }{\partial t} + \eta^i \frac{\partial }{\partial q^i} + \eta^{i \; [1]} \frac{\partial}{\partial \dot{q}^i} = \xi \frac{\partial }{\partial t} + \eta^i \frac{\partial }{\partial q^i} + (\dot{\eta}^i - \dot{q}^i \dot{\xi}) \frac{\partial}{\partial \dot{q}^i} \;.
\label{prolungamento vettore noether}
\end{equation}
It is called the \emph{first prolongation of Noether's vector}. We assume that our Lagrangian is not dependent on higher order derivatives and hence it is not necessary to calculate the transformation of $\ddot{q}^i$; nevertheless, it is possible to further extend the Noether vector to the $n$-prolongation as follows:
\begin{equation}
X^{[n]} = \xi \frac{\partial }{\partial t} + \eta^i \frac{\partial }{\partial q^i} + \eta^{i \; [1]} \frac{\partial}{\partial \dot{q}^i} + ... + \eta^{i \; [n]} \frac{\partial}{\partial \frac{d^n q^i}{dt^n}}
\end{equation}
here, it is 
\begin{equation}
\eta^{i \; [n]} = \frac{d \eta^{i \; [n-1]}}{dt} - \dot{\xi} \frac{d^n q^i}{dt^n} \;.
\end{equation}
Let us  show that if the coordinates transformation \eqref{trasformazione coordinate} leaves the equations of motion invariant, then the system satisfies the Noether identity
\begin{equation}
X^{[1]} \Lagr + \dot{\xi} \Lagr = \dot{g}(t,q^i) \;, \label{Teorema} 
\end{equation}
where $g$ is a generic function depending on coordinates and time. In order to prove the condition \eqref{Teorema}, we recall that the Euler-Lagrange equations are  \emph{invariant} if the following condition holds:
\begin{equation}
\frac{d\overline{t}}{dt} \overline{\Lagr} = \Lagr + \epsilon \dot{g}  \;.
\label{relazione di partenza}
\end{equation}
Deriving Eq. \eqref{relazione di partenza} with respect to $\epsilon$ and then setting  $\epsilon = 0$, we obtain:
\begin{equation}
\frac{d\overline{t}}{dt}  \frac{\partial \overline{\Lagr}}{\partial \epsilon} + \overline{\Lagr}\frac{\partial}{\partial \epsilon} \left(\frac{d\overline{t}}{dt}\right) = \dot{g} \;.
\label{relazione derivata}
\end{equation}
We can observe that $\displaystyle \frac{d\overline{t}}{dt} = \frac{\partial \overline{t}}{\partial t} + \frac{\partial \overline{t}}{\partial q^i} \dot{q}^i = 1 + \epsilon \frac{\partial \xi}{\partial q^i} \dot{q}^i$ that for $\epsilon = 0$, it  is equal to 1. Furthermore, assuming that it is possible to swap the order of  derivatives, we have that $\displaystyle \frac{\partial}{\partial \epsilon} \left(\frac{d\overline{t}}{dt}\right) = \frac{d}{dt}\frac{\partial \overline{t}}{\partial \epsilon} = \dot{\xi}$. Replacing these results into \eqref{relazione derivata} we obtain:
\begin{equation}
\xi \frac{\partial \overline{\Lagr}}{\partial t} + \eta^i \frac{\partial \overline{\Lagr}}{\partial \overline{q}^i} + \eta^{[1]\; i} \frac{\partial \overline{\Lagr}}{\partial \dot{\overline{q}}^i} + \dot{\xi} \overline{\Lagr} = \dot{g} \;,
\end{equation}
which is nothing else but \eqref{Teorema}. From this,  it  follows that systems satisfying the condition \eqref{Teorema} lead to the conserved quantity 
\begin{equation}
I(t,q^i,\dot{q}^i) = 	\displaystyle \xi \left(\dot{q}^i \frac{\partial \Lagr}{\partial \dot{q}^i} - \Lagr \right) - \eta^i \frac{\partial \Lagr}{\partial \dot{q}^i} + g(t,q^i) \;,
\end{equation}
which is a first integral of motion. For other techniques to integrate dynamical systems useful for cosmology see also \cite{Kamen1,Kamen2,Kamen3}.
\section{Noether symmetries in $N$-dimensional $f(\G)$ Cosmology}
\label{soluz}
Let us now apply the first prolongation of Noether vector to the Lagrangian \eqref{lagrangiana nD} whose generator, in our minisuperspace,  takes the form:
\begin{equation}
X = \xi(a,\G,t) \partial_t + \alpha(a,\G,t) \partial_a + \beta(a,\G,t) \partial_\G\,.
\end{equation} 
In order to find symmetries, we apply the Noether identity and  set  terms with derivative powers of  $a$ and $\G$ equal to zero. Therefore, the application of \eqref{prolungamento vettore noether} to \eqref{lagrangiana nD} gives  a system of four differential equations plus the constraints on the infinitesimal generators $\alpha, \beta, \xi$. It reads:
\begin{equation}
\begin{cases}
\displaystyle (n-1) a^2 \alpha (f- \G f') - 4 p(n) \dot{a}^2 \dot{\G} f''\partial_t \alpha - 
 a^3 \left[\beta f'' - (f - \G f') \partial_t \xi \right] = 0
\\
\displaystyle (n-4) \left[f' \partial_t \alpha + a f'' \partial_t \beta \right]= 0
\\
\displaystyle  (n-4 ) \alpha f'' + a \beta f''' + a f'' \left( 3 \partial_a \alpha + \partial_\G \beta - 3 \partial_t \xi \right) = 0
\\
\displaystyle (n-4)(n-5) \alpha f' + (n-4) a \beta f'' - (n-4 ) a f' \left(3 \partial_t \xi - 4 \partial_a \alpha \right) + 4 a^2 f'' \partial_a \beta = 0\,,
\label{System}
\end{cases}
\end{equation}
with 
$ \xi = \xi(t)\,,  \alpha = \alpha(a,t)\,,  g = g_0\,.$ Here,  we neglect \emph{a priori} the possibility $p(n) = 0$. Only three solutions satisfy the whole system; all of them provides the same dependence of the infinitesimal generator on the variables, namely
\begin{equation}
\alpha = \alpha_0 a\,, \quad \beta = \beta_0 \G\,, \quad \xi = \xi_0 t + \xi_1,
\end{equation}
but with different values of the constants $\alpha_0, \beta_0, \xi_0$. The final solutions with the corresponding infinitesimal generators are:
\begin{eqnarray}
&&1): \,\, \alpha = \alpha_0 a\,, \quad \beta = \beta_0 \G\,, \quad \xi = \alpha_0 \left(\frac{n-1}{3}\right) t + \xi_1\,, \qquad f(\G) = f_0 \G \nonumber
\\
&&2): \,\, \alpha = \alpha_0 a\,, \quad \beta = - 4 \xi_0 \G\,, \quad \xi = \xi_0 t + \xi_1\,, \qquad f(\G) = \frac{ 4 f_0 \xi_0}{\alpha_0(n-1) + \xi_0} \G^{\frac{\alpha_0(n-1) + \xi_0}{4 \xi_0}} \nonumber 
\\
&&3): \,\, \alpha = 0\,, \quad \beta = \beta_0 \G\,, \quad \xi =0\,, \qquad f(\G) = f_0 \G + f_1,
\label{soluznoeth}
\end{eqnarray}
where the exponent of the second function must be different from 1. The first and the third solution are non-trivial only in more than 4 dimensions, while the second provides contributions to the equations of motion even for $n=4$. Without loss of generality, in order to find the dynamics of the scale factor, we choose the function $f(\G) = f_0 \G^k$, where we define
\begin{equation}
\frac{\alpha_0(n-1) + \xi_0}{4 \xi_0} = k
\end{equation}
and we incorporated the coefficient of $\G^k$ into $f_0$. Here $k\in \mathbb{R}$. In this way, the point-like Lagrangian can be written as
\begin{equation}
\Lagr = -\frac{1}{3}a^{n-5} \G^{k-2} \left[3 (k-1) a^4 \G^2 + k (n-4) \G p(n) \dot{a}^4 + 4 k (k-1)a p(n) \dot{a}^3\dot{\G} \right].
\end{equation}
The Euler-Lagrange Eqs. \eqref{EL3} can now be exactly solved providing the following solutions:
\begin{eqnarray}
&& a(t) = a_0 e^{q t}\,, \qquad \G(t) = n p(n) \, q^4 \,,\qquad k= \frac{n}{4}
\\
&& a(t) = a_0 t^{- 4\frac{(k-1)\left(4k- 1\right)}{4k -n}}\,, \qquad \G(t) = \frac{256 \left[(k-1)\left(4k- 1 \right)\right]^3 \left[4 + n(4k-5) \right] p(n)}{(nt - 4kt)^4},
\label{soluzcosmo}
\end{eqnarray}
with $q$ constant. It is worth noticing that the de-Sitter-like expansion only holds in more that  4 dimensions, unlike the power-law solutions which is valid even for $n=4$. However, the $n=4$ case deserves a separate treatment. In next sections we will  focus on 4-dimensions in presence of matter. It is interesting to observe that the function containing a linear GB term leads to a solution with several free parameters which should be fixed out by experimental observations. It provides a vacuum exponential acceleration. 

\section{$f(\G)$ Cosmology in 4-Dimensions}
Let us now discuss specifically the four-dimensional case; in particular,  we will derive the  Noether symmetries  coming from the 4-dimensional Lagrangian and the related cosmological  solutions in presence of matter. After,  we will also consider  the case of GB term non-minimally coupled with a scalar field. We introduce the matter Lagrangian through the choice $\Lagr_m = \rho_0 a^{-3 w}$, where $w$ represents the ratio between pressure and density $p = w \, \rho$, that is the Equation of State of a perfect fluid. For $w = 0$, we have dust matter,  for $w = \frac{1}{3}$, we have radiation. The case $w = -1$, in turn, corresponds to the cosmological constant. Therefore, being $p(4) = 6$, the Lagrangian \eqref{lagrangiana nD}, in 4-dimensions,  is
\begin{equation}
\Lagr^{(4)} = a^3[ f(\G) - \G f'(\G)] - 8 \dot{a}^3 f''(\mathcal{G}) \dot{\mathcal{G}} + \rho_0 a^{-3 w} \;.
\label{lagrangiana}
\end{equation}
The Euler-Lagrange equations of the above Lagrangian read as:
\begin{equation}
\begin{cases}
\displaystyle \frac{d}{dt}\frac{\partial \Lagr}{\partial \dot{\G}} = \frac{\partial \Lagr}{\partial \G} \;\;\; \to \;\;\; \G = 24 \frac{\dot{a}^2 \ddot{a}}{a^3}
\\
\\
\displaystyle \frac{d}{dt}\frac{\partial \Lagr}{\partial \dot{a}} = \frac{\partial \Lagr}{\partial a} \;\;\; \to \;\;\; a^2[f(\G) - \G f'(\G)] + 16 \dot{a} \ddot{a} \dot{\G} f''(\G) + 8\dot{a}^2[f''(\G) \ddot{\G} + f'''(\G) \dot{\G}^2]  + 3 \rho_0 w a^{-3w-1} \;.
\end{cases}
\end{equation}
The first equation is the Lagrange multiplier in 4-dimensions.   Finally we  have to take into account the energy condition 
\begin{equation}
E_\Lagr = \dot{a} \frac{\partial \Lagr}{\partial \dot{a}} + \dot{\G} \frac{\partial \Lagr}{\partial \dot{\G}} - \Lagr = 0 \;,
\end{equation} 
which gives 
\begin{equation}
a^3[f(\G) - \G f'(\G)] + 24 \dot{a}^3 f''(\G) \dot{\G} + \rho_0 a^{-3 w} = 0 \;.
\end{equation}
Applying the Noether condition \eqref{Teorema} to the Lagrangian \eqref{lagrangiana}, we get a system of two differential equations, since the second equation appearing into \eqref{System} canceled out for $n=4$ and the fourth trivially reduces to $\partial_a \beta =0$. The system  takes the form:
\begin{equation}
\begin{cases}
\displaystyle 3 \al a^2 [ f(\G) - \G f'(\G) - w \rho_0 a^{-3(w+1)}] - \be a^3 \G f''(\G) + \partial_t \xi a^3 [ f(\G) - \G f'(\G)] = 0
\\
\displaystyle 3 \partial_a \alpha 	 f''(\G)  + \be f'''(\G) - 3 \partial_t \xi\; f''(\G) + \partial_\G \be \; f''(\G) = 0
\\
\displaystyle \xi = \xi(t)\,, \quad \alpha = \alpha(a)\,, \quad \beta = \beta(\G) \quad g = g_0.
\end{cases}
\end{equation}
The presence of the matter Lagrangian does not cause any changes in the system resolution, so that the function and the infinitesimal generator turn out to be the $n=4$ case  of  \eqref{soluznoeth} assigning the Noether vector,  namely
\begin{equation}
\begin{cases}
\al = \al_0 a\,, \quad  \xi_0 t + \xi_1\,, \quad  \be = -4 \xi_0 \G\,, \quad g = g_0
\\
\displaystyle f(\G) = \frac{f_0 \G^k}{k}\,, \qquad k \neq 1 \;.
\end{cases}
\end{equation}
By using the above solutions and incorporating the constant $k$ into $f_0$, we can rewrite the point-like Lagrangian as
\begin{equation}
\Lagr = - f_0 (k-1) \G^k a^3 - 8 f_0 k (k-1) \G^{k-2} \dot{a}^3 \dot{\G} - \rho_0 a^{-3w} \;.
\label{Lagrangiana finale}
\end{equation}
Euler-Lagrange equations and energy condition coming from \eqref{Lagrangiana finale} lead to the system
\begin{equation}
\begin{cases}
\displaystyle \G = 24 \frac{\dot{a}^2 \ddot{a}}{a^3}
\\
\\
\displaystyle -8 f_0 k (k-1)(k-2) \dot{a}^2 \G^{k-3} \dot{\G}^2 - 8 f_0 k(k-1) \dot{a} \G^{k-2} \left(2 \ddot{a} \dot{\G} + \dot{a} \ddot{\G}\right)+f_0 (k-1) a^2 \G^k -\rho_0 w a^{-3 w-1} = 0
\\
\\
\displaystyle -24 f_0 k (k-1) \dot{a}^3 \G^{k-2} \dot{\G} + f_0 (k-1) a^3 \G^k +\rho_0 a^{-3w}= 0 .
\end{cases}
\label{EL2}
\end{equation}
There are two kinds of solutions of the above system; the first can be obtained neglecting the matter Lagrangian. In this case, when geometric contributions are  greater than matter ones, the only solution reads 
\begin{equation}
a(t) = a_0 t^{1-4k} \;\;\;\;\;\; \G(t) = -96 k(1-4k)^3 t^{-4} \equiv \G_0 t^{-4} \;,
\label{soluzioni cosmo}
\end{equation}
which is a  power-law expansion and, as expected, it is contained into \eqref{soluzcosmo}. Without neglecting $\Lagr_m$, we find another set of solutions, namely:
\begin{equation}
\begin{cases}
a(t) = a_0 t^{1-4k} \;\;\;\;\;\; \G(t) = -96 k(1-4k)^3 t^{-4} \equiv \G_0 t^{-4} \;, \,\,\,\,\,\,\,\, w = 0
\\
\\
a(t) = e^{n t} \,\,\,\,\,\,\, \G(t) = 24 m^4  \,\,\,\,\,\,\, w = -1 .
\end{cases}
\end{equation}
The former is the solution for dust matter, while in the latter  the matter plays the role of cosmological constant. 
Nevertheless, from Eq. \eqref{soluzioni cosmo}, we can distinguish the cosmological  eras depending on the geometrical contributions even in vacuum:
\begin{equation}
\begin{split}
&k = \frac{1}{8} \; \to \; a(t) \sim t^{\frac{1}{2}} \; \;\; \G = - \frac{3}{2}  t^{-4} \;\; \to \; \mathrm{Radiation}
\\
&k = \frac{1}{6}\; \to a(t)  \sim t^{\frac{1}{3}} \; \;\; \G = - \frac{16}{27} t^{-4} \;\; \to \; \mathrm{Stiff \; matter} 
\\
&k = \frac{1}{12} \; \to \; a(t) \sim t^{\frac{2}{3}} \; \;\; \G = - \frac{64}{27} t^{-4} \;\; \to \; \mathrm{Dust \; matter}  \;.
\end{split}
\end{equation}
Cosmological solutions \eqref{soluzioni cosmo}  are, therefore,  in agreement with the FRW solutions of GR but are recovered without imposing the Ricci scalar in the gravitational action. It is worth noticing  that in all cases the Gauss-Bonnet term turns out to be negative, so that the function $f(\G) = f_0 \G^k$ may lead to some problems for  fractional even values of $k$. To avoid these kind of singularities, we want to stress that the function $f(\G)$ is still a solution of Noether's system even including the modulus of the Gauss-Bonnet term, i.e.  $f(\G) = f_0 |\G|^k$. The same happens in several other modified theories; for example, in $f(R)$ gravity, the Noether approach provides the solution $f(R) \sim R^{3/2}$ \cite{Capozziello:2008ch}, whose time power-law solution $ a(t) \sim t^p$ leads to a complex function for $p<0$ and $p > 1/2$. Hence, without loss of generality and in agreement with Noether's approach, we can always require the function into the action to be positive. However, as shown in \cite{Prado} for $f(R)\sim |R|^{3/2}$, some exact solutions can imply transitions from decelerated/accelerated behaviors, that is dust/dark energy behaviors according to the values of solution parameters. In the present case, however, we are discussing only exact solutions emerging from Noether's symmetries where there is no change of concavity in the evolution of the scale factor and then no transitions from decelerated to accelerated behaviors and viceversa.

\subsection{Brans-Dicke coupling}
\label{branse}
Dynamics can be improved by coupling the GB function  coming from the existence of Noether symmetries  with  a scalar field $\phi$. In such a way the scalar-tensor action reads:
\begin{equation}
S = \int \sqrt{-g}\left[ \phi \G^k + \frac{\omega(\phi)}{\phi} \partial^\mu \phi \partial_\mu \phi + V(\phi)\right] d^4 x \;.
\end{equation}
In this perspective,  we can consider the simplest form of scalar-tensor theories, with zero potential and $\omega(\phi) \equiv \omega$, that is a Brans-Dicke theory coupled  with  GB geometry. The action becomes:
\begin{equation}
S = \int \sqrt{-g}\left[ \phi \G^k + \frac{\omega}{\phi} \partial^\mu \phi \partial_\mu \phi \right] d^4 x \;.
\end{equation}
In this case, the field equations can be written as:
\begin{eqnarray}
&&2k R \nabla_\mu \nabla_\nu \G^{k-1} - 2 k g_{\mu \nu} R \Box \G^{k-1} - 4 k R^\lambda_\mu \nabla_\lambda \nabla_\nu \G^{k-1} + 4 k R_{\mu \nu} \Box \G^{k-1} \nonumber
\\
&&+ 4 k g_{\mu \nu} R^{p \sigma} \nabla_p \nabla_\sigma \G^{k-1} + 4 k R_{\mu \nu p \sigma} \nabla^p \nabla^\sigma \G^{k-1} + \frac{1}{2} (1-k) g_{\mu \nu}[\G^{k}]  + \nonumber
\\
&& -{\frac  {\omega }{\phi ^{2}}}(\partial _{\mu}\phi \partial _{\nu}\phi -{\frac  {1}{2}}g_{{\mu \nu}}\partial _{p}\phi \partial ^{p}\phi )-{\frac  {1}{\phi }}(\nabla_{\mu}\nabla_{\nu}\phi -g_{{\mu \nu}}\Box \phi )= 0 \;.
\end{eqnarray} 
As above,  the corresponding point-like Lagrangian is 
\begin{equation}
\Lagr =  k(1-k) \phi(t) \dot{a}(t)^3 \G(t)^{k-2} G'(t) -k \dot{a}(t)^3 \G(t)^{k-1} \dot{\phi}(t)+ (1-k) a(t)^3 \phi(t) \G(t)^k - \frac{\omega a(t)^3 \dot{\phi}(t)^2}{\phi(t)} \;.
\label{La brans}
\end{equation}
Euler-Lagrange equations and the energy condition for \eqref{La brans} are 
\begin{eqnarray}
\frac{d}{dt} \frac{\partial \Lagr}{\partial \dot{a}} = \frac{\partial \Lagr}{\partial a} \;\; \to \;\; && \frac{6 \omega a^2 \dot{a} \dot{\phi}}{\phi}+ 3 k \dot{a}^2 \ddot{a} \G^{k-1}+ a^3 \left[(k-1) \G^k- \frac{\omega \left(\dot{\phi}^2- 2 \phi \ddot{p}\right)}{\phi^2}\right] =0\nonumber
\\
\frac{d}{dt} \frac{\partial \Lagr}{\partial \dot{\phi}} = \frac{\partial \Lagr}{\partial \phi} \;\; \to \;\; && \frac{3 a^2 \left[(k-1) \phi^2 \G^k+ \omega \dot{\phi}^2\right]}{\phi}-3 k \dot{a} \G^{k-3}
   \left\{\G \left[2 \G \ddot{a} \dot{\phi}+\dot{a} \left(2 (k-1) \dot{G} \dot{\phi}+\G
 \ddot{\phi}\right)\right]+\right. \nonumber 
 \\
 &&\left. (k-1) \phi \left[2 \G \ddot{a} \dot{\G}+\dot{a} \left(\G
   \ddot{G}+(k-2) \dot{\G}^2\right)\right]\right\} = 0\nonumber
\\
\frac{d}{dt} \frac{\partial \Lagr}{\partial \dot{\G}} = \frac{\partial \Lagr}{\partial \G} \;\; \to \;\; && \G = \frac{\dot{a}^2 \ddot{a}}{a^3}\nonumber
\\
\dot{q^i} \frac{\partial \Lagr}{\partial \dot{q^i}} - \Lagr = 0 \;\; \to \;\; && \nonumber \frac{a^3 \left[(k-1) \phi^2 \G^k- \omega \dot{\phi}^2\right]}{\phi}- k \dot{a}^3 \G^{k-2}
   \left[2 (k-1) \phi \dot{\G}+3 \G \dot{\phi}\right] = 0 \;
\end{eqnarray}
and can be easily solved giving the de Sitter solution
\begin{equation}
a(t) = a_0 e^{nt}\,, \;\;\;\;\; \phi(t) = \exp \left\{ \left[n^{2k} \frac{\sqrt{\omega(k-1)}}{\omega}\right] t \right\}\,,\;\;\;\; \G(t) = 24 n^4 \;.
\end{equation}
By coupling the scalar field to GB term, accelerated expansion is recovered even if the scalar-field self-interaction potential $V(\phi)$ is not present. Einstein's gravity,  even in this case, is recovered for $k= \frac{1}{2}$; contributions to GB term coming from \textbf{Riem}$^2$ and \textbf{Ricci}$^2$, in some cosmological context, are comparable to $R^2$, so that $\sqrt{\G} = \sqrt{R^2 - 4 \textbf{Ricci}^2 + \textbf{Riem}^2 } \sim \sqrt{R^2}$. It means that, in some epochs, $R^2$ and $\G$ are dynamically equivalent up to a constant term.  In fact, considering   power-law solutions of the form $a(t) \sim t^p$, we have
\begin{equation}
\G = 24 \frac{\ddot{a} \dot{a}^2}{a^3} = \frac{24 p^3(p-1)}{t^4} \qquad R = -6 \left(\frac{\ddot{a}}{a} + \frac{\dot{a}^2}{a^2} \right) = \frac{6p (2p-1)}{t^2} 
\end{equation} 
so that it is clear that $\G \sim R^2$ being $p$ a number. The same holds for exponential solutions, where both $R$ and $\G$ are constants and independent of time. Therefore, $\G$ and $R^2$ can be considered dynamically equivalent on the solutions (up to a constant factor) if homogeneity and isotropy hold. A more general case is the one concerning the sum of different powers of $\G$ that can be easily reduced to  to $R+f(\G)$ or $f(R)+f(\G)$.
\subsection{The case $\G^n + \G^k$}
\label{sum}
In this section we deal with  the case of a function made of a sum of powers of the GB term, in 
 4-dimensions. Even though it is not directly a solution of the Noether system and it does not contain symmetries in this metric, it could be very relevant for several reasons. We mainly want to stress that in cosmology, in some epochs, GR is recovered with the choice $f(\G) = \sqrt{\G}$. In spherical symmetry, something similar happens for different $k$, as shown in \cite{Baj}, where  the Noether approach is applied to a pure spherically symmetric GB theory. A function like
\begin{equation}
f(\G) = f_0 \G^{\frac{1}{2}} + f_1 \G^{n} \;,
\end{equation}
can easily be compared to the case $f(R,\G)=R+f(\G)$,  often discussed in literature in view to recover GR in suitable limits \cite{Odintsov3,Odintsov4}. 

We generalize the concept by considering the function $f(\G) = f_0 \G^n + f_1 \G^k$; the Lagrangian is a particular case of \eqref{lagrangiana} and it reads:
\begin{equation}
\Lagr = -a^3 \left[f_0(n-1) \G^n + f_1 (k-1) \G^k \right] - 8\left[f_0 n(n-1) \G^{n-2} + f_1 k(k-1) \G^{k-2} \right] \dot{a}^3 \dot{\G} \;.
\end{equation}
The Euler-Lagrange equations and  the energy condition  are:
\begin{eqnarray}
\displaystyle \frac{d}{dt}\frac{\partial \Lagr}{\partial \dot{\G}} = \frac{\partial \Lagr}{\partial \G} \;\;\; \to \;\;\; && \G = 24 \frac{\dot{a}^2 \ddot{a}}{a^3} \nonumber
\\
\displaystyle \frac{d}{dt}\frac{\partial \Lagr}{\partial \dot{a}} = \frac{\partial \Lagr}{\partial a} \;\;\; \to \;\;\; && 3 a^2 \left[f_0 (n-1) \G^n+ f_1 (k-1) G^k\right]-24 \dot{a} \left[f_0 n \left(n^2-3 n+2\right) \dot{a} \G^{n-3} \dot{\G}^2 +\right. \nonumber
\\
&& +f_1 k \left(k^2-3
   k+2\right)\dot{a} \G^{k-3} \dot{\G}^2+f_0  n (n-1) \G^{n-2} \left(2 \ddot{a}
   \dot{\G}+\dot{a} \ddot{G}\right)+ \nonumber 
   \\
   && \left.+ f_1 k (k-1)  \G^{k-2}  \left(2 \ddot{a} \dot{\G}+\dot{a} \ddot{G}\right)\right] \nonumber
\\
\dot{a} \frac{\partial \Lagr}{\partial \dot{a}} + \dot{\G} \frac{\partial \Lagr}{\partial \dot{\G}} - \Lagr = 0 \;\;\; \to \;\;\;  &&  a^3 \left[f_0 (n-1) \G^n +f_1 (k-1) \G^k\right]-	24 \dot{a}^3 \dot{\G} \left[
f_0 n (n-1)  \G^{n-2}+f_1 k (k-1) \G^{k-2}\right].
\end{eqnarray}
The system  admits the following de Sitter solution:
\begin{equation}
a(t) = a_0 e^{mt} \;\;\; \G(t) = 24 m^4 \;\;\; \text{with} \;\; m = \left[-24^{n - k}\; \frac{f_0}{f_1} \left(\frac{n-1}{k-1}\right) \right]^{\frac{1}{4(k-n)}}\,.
\end{equation}
This means that dark energy \cite{Odintsov3} and inflation \cite{Paolella} can be easily recovered in this framework.

\section{Quantum Cosmology and the  Wave Function of the Universe}
\label{Hamiltonian}

The above considerations allow to develop also quantum cosmology for the minisuperspace $T\mathcal{Q}\equiv\{a,\dot{a}, \G, \dot{\G}\}$.
Starting from Lagrangian \eqref{Lagrangiana finale} we can calculate the related Hamiltonian as a function of momenta:
\begin{equation}
\mathcal{H} = \frac{f_0}{k} \G^k a^3 + \pi_a \left(-\frac{\pi_\G}{8f_0} \G^{2-k} \right)^{\frac{1}{3}} \;,
\label{Hamiltoniana finale}
\end{equation}
where $\pi_a=\frac{\partial {\cal L}}{\partial \dot {a}}$ and $\pi_{\G}=\frac{\partial {\cal L}}{\partial \dot {\G}}$ according to the Legendre transformations. 

Thanks to the Noether symmetries, we can insert into \eqref{Hamiltoniana finale},  a cyclic variable which allows to fully quantize the theory.  From Eqs. \eqref{soluzioni cosmo},  it is easy to see that the quantity $\displaystyle \frac{\dot{a}}{\G^k}$ is a constant of motion. Immediately, it is
\begin{equation}
\frac{\dot{a}^3}{\G^{3k}} = \Sigma_0 \;,
\end{equation}
and then we can rewrite $\pi_\G$ as 
\begin{equation}
\pi_\G = -8 f_0 \Sigma_0 \G^{4k-2} \;.
\end{equation}
Replacing this result into \eqref{Hamiltoniana finale},  we can write the Hamiltonian in the simpler form:
\begin{equation}
\mathcal{H} = \frac{f_0}{k} \G^k a^3 + \pi_a \left(\Sigma_0 \G^{3k} \right)^{\frac{1}{3}} \;.
\end{equation}
Now, thanks to the quantization rules coming from the Arnowitt-Deser-Misner (ADM)  formalism \cite{Misner, ADM}, we can define the operators
\begin{equation}
\begin{cases}
\displaystyle \pi_\G = -i \frac{\partial}{\partial \G}
\\
\\
\displaystyle \pi_a = - i \frac{\partial}{\partial a}
\\
\\
\displaystyle \mathcal{H} \psi = 0 \;.
\end{cases}
\end{equation}
The third equation is the so-called Wheeler-de Witt Equation and  $\psi$ is the Wave Function of the Universe \cite{Misner, ADM,DeWitt:1967yk, DeWitt:1967ub, Thiemann:2007zz}.

From  the first equations, being  $\displaystyle \pi_\G = -8 f_0 \Sigma_0 \G^{4k-2}$, the quantity $\displaystyle \pi_\G \G^{2-4k}$ is a constant of motion. More precisely the quantized equation of momentum can be written as:
\begin{equation}
i \frac{\partial}{\partial \G} \psi(a,\G) = 8 f_0 \Sigma_0 \G^{4k-2} \psi(a,\G),
\end{equation}
so that we get the system 
\begin{equation}
\begin{cases}
\pi_\G \psi = -i \frac{\partial}{\partial \G} \psi \;\;\;\;\; \to \;\;\;\;\; \psi(a,\G) = A(a) \; \displaystyle \exp\left\{i \; \frac{8 f_0 \Sigma_0 \G^{4k-1}}{1-4k}\right\}
\\
\mathcal{H} \psi = 0 \;\;\;\; \to \;\;\;\; \displaystyle\frac{f_0}{k} \left(\Sigma_0 \right)^{-\frac{1}{3}} a^3 A(a) - i \displaystyle\frac{\partial A(a)}{\partial a} = 0 \;.
\end{cases}
\end{equation}
The latter equation has the  solution:
\begin{equation}
A(a) = A_0 \exp\left\{-\frac{i}{4} \frac{f_0}{k} \left(\Sigma_0 \right)^{-\frac{1}{3}} a^4 \right\}
\end{equation}
and hence finally the Wave Function of the Universe is
\begin{equation}
\psi(a,\G) = \psi_0 \exp\left\{i\left[-\frac{f_0}{4k} \left(\Sigma_0 \right)^{-\frac{1}{3}} a^4 + \frac{8 f_0 \Sigma_0 \G^{4k-1}}{1-4k} \right]\right\} \;. \label{WF}
\end{equation}
According to the  Hartle criterion \cite{Hartle2,Hartle1}, an  oscillating Wave Function means correlations among variables and then the possibility to find classical trajectories (i.e. observable universes). In fact, considering  the WKB approximation, it is  $\psi(a,\G) \sim e^{iS}$ (where $S$ is the action), we have, from   \eqref{WF}, 
\begin{equation}
S = -\frac{f_0}{4k} \left(\Sigma_0 \right)^{-\frac{1}{3}} a^4 + \frac{8 f_0 \Sigma_0}{1-4k} \G^{4k-1}
\end{equation}
and, after some trivial calculations, we notice that Hamilton-Jacobi equation with respect to the scale factor provides the third equations of motion in \eqref{EL2}:
\begin{equation}
\displaystyle \frac{\partial S}{\partial a} = \pi_a \;\;\;\; \to \;\;\;\; \G^k a^3 = 24 \G^{k-2} \dot{a}^3 \dot{\G} \;.
\end{equation}
The second Hamilton-Jacobi equation $\displaystyle \left(\frac{\partial S}{\partial \G} = \pi_\G \right)$ instead, is nothing but the identity $\pi_\G = \Sigma_0$ which can be recast into the second equation of motion of \eqref{EL2}. In this sense, classical trajectories, and then observable universes, are recovered.
As reported in \cite{Capozz5,Lambiase}, oscillatory behaviors of the Wave Function of the Universe are related to  conserved quantities coming from Noether symmetries. If the number of symmetries is equal to the variables of minisuperspace, the dynamical system is fully integrable and the Wave Function fully oscillating. As a consequence, we can state that Noether symmetries select observable universes.

\section{Discussion and Conclusions}
\label{concl}
In this paper, we discussed $f(\G)$ cosmology via the Noether Symmetry Approach. The main results  are that the  existence   of symmetries selects a power-law  form of $ \displaystyle f(\G) = f_0 \G^k$ and,   in 4-dimensions,   with the further constraint $k \neq 1$, we can obtain interesting dynamics.  Furthermore,  we can observe that the case $k=1$  is not allowed, in agreement with the fact that $\displaystyle S = \int_{\cal{M}} \G d^4x = \chi(\cal{M})$, being $\chi(\cal{M})$ the Euler characteristic. Moreover, taking into account the definition of the Gauss Bonnet invariant $\G$  and considering that,  in FRW cosmology $R_{\mu \nu} R^{\mu \nu}$, $R^{\mu \nu p \sigma} R_{\mu \nu p \sigma} \ll R^2$, for $k = 1/2$ we can recover Einstein's gravity. In other words,   GR can be seen as a particular case  of  $f(\G)$ theory without asking for the corrected theory $R+f(\G)$. 
In this framework, it is possible to obtain both exponential and power-law cosmological solutions also in presence of standard matter. The former can be recovered only in 5-dimensions or more, while the latter can be found  even in 4-dimensions. In $4$-dimensions,    de Sitter solutions are possible only adding an extra  term $\Lagr_m \sim e^{-3w}$ with $w=-1$. Nevertheless,  coupling  $\G^k$ to a scalar field $\phi$, de Sitter exponential law is immediately recovered also in  in vacuum.  

Furthermore,  we analyzed the  sum
$f(\G) = f_0 \G^n + f_1 \G^k$ which, according to the above considerations,  naturally can give  $f(R,\G)=R+f(\G)$.  Also in this case, we found exact solutions. 

Finally, we discussed the quantum cosmology for the minisuperspace related to the variables $a$ and $\G$. Also in this case, symmetries have a key role for the interpretation of the Wave Function of the Universe.
They allow to find out oscillatory behaviors and then the possibility to apply the Hartle criterion, which states that oscillations mean correlations between variables and then the possibility to achieve classical trajectories, that is  observables universes.

As a concluding remark, considering extended Gauss-Bonnet cosmology can result useful from several points of view, in particular, for avoiding ghost modes  \cite{Capozziello:2019klx} and other pathologies present in GR and in other modified gravity theories.  Beside this fact, it seems a natural approach towards quantum fields in curved spaces and, finally, towards quantum gravity \cite{Buchbinder}.

\section*{Acknowledgments}
The Authors  acknowledge the support of  {\it Istituto Nazionale di Fisica Nucleare} (INFN) ({\it iniziative specifiche} MOONLIGHT2 and QGSKY). This paper is based upon work from COST action CA15117 (CANTATA), supported by COST (European Cooperation in Science and Technology).


\end{document}